\documentclass[12pt]{article}
\usepackage{epsfig}

\parskip=\medskipamount 
plus.3em minus.5em \abovedisplayshortskip=.5em plus.2em minus.4em
\renewcommand{\thesection}{\arabic{section}}

\catcode`@=11

\@addtoreset{equation}{section} \@addtoreset{equation}{subsection}
\def\theequation{\ifnum\value{section}=0 \arabic{equation}\ignorespaces
\else \ifnum\value{section}=-1 A.\arabic{equation}\ignorespaces
\else \ifnum\value{subsection}=0
\thesection.\arabic{equation}\ignorespaces \else
\thesection.\arabic{subsection}.\arabic{equation}\ignorespaces
                             \fi
                        \fi
                   \fi}

{\catcode`\'=\active \def'{{}^\bgroup\prim@s}}

\catcode`@=12

\newcommand{\bq}{\begin{equation}}
\newcommand{\be}{\begin{equation}}
\newcommand{\fq}{\end{equation}}
\newcommand{\ee}{\end{equation}}
\newcommand{\bqr}{\begin{eqnarray}}
\newcommand{\beqs}{\begin{eqnarray}}
\newcommand{\fqr}{\end{eqnarray}}
\newcommand{\eeqs}{\end{eqnarray}}

\newcommand{\rf}[1]{(\ref{#1})}

\def\bop#1{\setbox0=\hbox{$#1M$}\mkern1.5mu
    \vbox{\hrule height0pt depth.04\ht0
    \hbox{\vrule width.04\ht0 height.9\ht0 \kern.9\ht0
    \vrule width.04\ht0}\hrule height.04\ht0}\mkern1.5mu}
\def\Box{{\mathpalette\bop{}}}                        

\begin{document}
\thispagestyle{empty}

\begin{flushright}
\begin{tabular}{l}
hep-th/0507186 \\
\end{tabular}
\end{flushright}

\vskip .6in
\begin{center}

{\bf  Computing K3 and CY n-fold Metrics}

\vskip .6in

{\bf Gordon Chalmers}
\\[5mm]

{e-mail: gordon@quartz.shango.com}

\vskip .5in minus .2in

{\bf Abstract}

\end{center}

The derivative expnsion in the context of IIB string scattering compactified 
on non-trivial K3 and other Calabi-Yau manifolds is formulated.  The scattering 
data in terms of automorphic functions can be inverted to find the these metrics.  
The solutions are parameterized by the moduli information, and the metrics may 
be found to any desired accuracy in derivatives.  Metric information to low orders 
in derivatives allows for a counting of curves inside the manifold; in addition, 
the coefficients of these exponential terms via D-brane wrappings are polynomials 
that may admit an invariant interpretation in cohomology.  An interesting case 
pertaining to M-theory compactifications is the collection of seven-dimensional 
$G_2$ manifolds; they can also be obtained when the moduli space degenerates into 
cases, such as a toroidal one or other limit in which modular functions on the 
space are known.\footnote{This work was written two years ago; the recipe without 
the explicit form of the scattering and metrics is given.}

\vfill\break 

\section{Introduction} 

Compact Ricci-flat manifolds were discovered in 1957, but to date there is moderate 
progress in computing their metric form away from the large volume region or in some 
degenerate limit.  Knowing the explicit metric would allow for a variety of 
computations: holomorphic curve counting, gravitational instanton contributions to 
string backgrounds, number theory results, and more.  In this work the derivative 
expansion applied to gravitational theories, together with S-duality, generate a 
recipe for computing the K\"ahler potential of the manifold to an arbitrary order in 
derivatives; formally the complete classical gravitational scattering in the K3 
background generates the K3 metric.  (Recent work in \cite{numericalK3} generated 
numerically partial metrics on certain K3 spaces.)

First, the method presented to computing the metrics is described; model dependencies 
vary the procedure slightly.  The bosonic truncation of the covariantized gravitational 
effective action has the form in Einstein frame,

\bqr 
{\cal S} = {1\over G_N^2} ~\int~ d^dx \sqrt{g} ~ \bigl[ R + \alpha_1 \ln\Box R^2 
+ \alpha_2 R^3 + \ldots \ , 
\label{prototypeN2} 
\fqr 
and whose variation generates the on-shell S-matrix evaluated in some background.  The 
coefficients parameterizing the effective action, in general, are computable through 
perturbative and non-perturbative methods.  In theories with non-perturbative duality 
structure such as IIB superstring theory with $\tau\leftrightarrow -{1\over\tau}$, the 
coefficients of the curvatures are easier to obtain.  The coefficients are obtainable 
via pseudo free-field x-space diagrams in the derivative expansion.\footnote{The work
in the context of various theories is described in \cite{Chalmers1}-\cite{Chalmers16}.}  
IIB compactified 
on K3, and other certain Calabi-Yau manifolds, allow the non-perturbative form to be 
deduced via perturbative methods.  The non-perturbative terms correspond to various 
D-branes (i.e. D-instantons, D1 and D3 branes) wrapping the internal cycles to be 
computed; they are connected to the perturbative terms through S-duality, similar to 
${\cal N}=4$ supersymmetric gauge theory \cite{Chalmers10} as the gauge theory instantons are 
derivable from the perturbative terms.  

The idea in obtaining the non-trivial metrics of K3, or other Calabi-Yau n-folds, 
involves first substituting $g_{i\bar j}=\partial_i\partial_{\bar j} \phi$ within 
the effective action.  The vanishing of the action density 

\bqr 
 {\cal S} [\phi] = 0  \qquad S= \int {\cal S}[\phi] 
\label{variation} 
\fqr 
should generate the K\"ahler potential; the kinetic term $\int R$ is a total derivative, 
and the total action in the quantum regime should vanish as there is a minimum to the 
action on the K\"ahler manifold.  Whether there is an integration by parts or not, there 
should be a minimum.  This variation is non-trivial due to both the presence of the logarithmic 
terms in the effective actions and the effective vertices which depend on the modular 
forms commanded by S- and T-duality.  This routine in principle generates the metric 
to an arbitrary order of accuracy, in addition to allowing the computation of all of the 
gravitational instantons wrapping the sub-cycles within the manifold (and also the 
number and type of cycles in the manifold).  

\section{Nonperturbative coefficients on CY n-folds} 

First, the IIB superstring derivative expansion in a flat-space background, as formulated 
in \cite{Chalmers16}, is reviewed.  The generating function of the scattering amplitude 
at two to four-point order is found from, 

\bqr 
{\cal S} = {1\over\alpha^4} ~\int ~ d^{10}x\sqrt{x} ~ \bigl[ R + R{1\over\Box}R + 
\alpha^3 \sum_{k=1}^\infty f_k(\tau,\bar\tau) (\alpha\Box)^k R^4 + {\rm non-analytic} 
\bigr] \ , 
\fqr 
with the derivatives acting genrerally within the components of the contracted $R^4$ 
terms (the tensor of the $R^4$ term is determined by maximal supersymmetry).  The 
non-analytic terms have the general expansion of logarithms mingling within the curvature 
terms; they may be computed via unitarity - the imaginary parts of amplitudes should be 
found via Cutkowsky rules.  The prefactors $f_k$ are modular functions and under S-duality 
transform under $\tau\rightarrow (a\tau+b)/(c\tau+d)$ with $ac-bd=-1$.  The remaining 
terms in the action are found from a supersymmetric extension of the action (for example, 
with a lightcone maximal superspace of \cite{Lightcone}).  

The modular functions at a given order $k$ are spanned by the ring of functions, 

\bqr  
\prod E_{s_i}^{(q_i,-q_i)} \ , \qquad \sum s_i = s, \qquad \sum q_i = 0 \ , 
\fqr 
and 

\bqr 
E_s^{(q,-q)} = \sum_{(m,n)\neq (0,0)} {\tau_2^s\over (m+n\tau)^{s+q} (m+n\bar\tau)^{s-q}} \ , 
\fqr 
with the sum ranging over coprime pairs of integers and $s=3/2+k/2$.  The ring element 
coefficients (including possible non-holomorphic $q_i\neq 0$ terms) are determined 
by sewing relations, after including the supersymmetric terms. 

The formulation previously discussed is valid when the vacuum of the IIB string is 
described by the coset SL(2,R)/U(1), the keyhole region in the complex region.  When 
the vacuum moduli space changes, a different set of automorphic functions enter into 
the description; for example, S-duality is expected for the IIB string compactified on 
a multi-torus, and the moduli space of the vacuum is described by the semi-direct 
product of the SL(2,R) action and the T-duality group of the torus $SO(10-d,10-d)/ 
SO(10-d)/SO(10-d)$, a member of the potentially affine duality group $E_{11-d,11-d}$. 

Next we examine the general graviton scattering on the K3 manifolds.  These manifolds 
have $57$ moduli, with a rather simple moduli space, $SO(3,19)/SO(19)/SO(3)$ [231-3-171 
=57].  The semi-direct product of S and T-duality forms the moduli (Teichmuller) space 
$SO(5,21)/SO(5)/SO(21)$; 
because the K3s moduli space is toroidal it is possible to formulate the automorphic 
functions taking values on it; we may also derive the string scattering and derive the 
explicit set of metrics.  

Consider the scattering of IIB on ${\rm T}^6\times$K3, with the base space 
being ${\rm K}3$.  The number of supersymmetries is naively $32$ components as found 
on the K3 base, with one covariantly conserved K3 spinor times the eight spinors arising 
from the $T^6$ torii; however, the theory has only ${\cal N}=4$ in $d=4$ with the grav 
multiplet arising from the K3 and the toroidal space generating a tower of ${\cal N}=4$
multiplets.  S-duality is expected to be present in theories containing ${\cal N}\geq 
16$ supersymmetries.    We will neglect all o fthe massless and massive modes entering 
from the higher dimensional space, i.e. the $T^6$, by treating the theory with only the 
K3 moduli space and neglecing these Kaluza-Klein modes in the sewing procedure; this 
allows us to examine the string on the reduced four-dimensional space-time.  Furthermore, 
the massive string modes are to be decouple, leaving only the massless theory containing 
the couplings dictated by the moduli of the K3 background; the latter is the Lagrangian 
description of the theory that we examine in the following.  Dropping the massive modes 
allows the integrals to be computed without detailed knowledge of the massive mode couplings; 
in principle, however, M-theory graviton scattering can be used to obtain the coefficients 
of the S-duality invariant graviton scattering without the massive mode technicalities, as 
described later.  

The gravitational string scattering on the K3 part of the K3$\times T^6$ is described by 

\bqr  
{\cal S}^{n-pt}_{\rm grav} = \sum_{k=1}^\infty \int d^4x \sqrt{g}~ {\cal O}_k \beta_k \ , 
\label{effective} 
\fqr 
similar to the flat ten-dimensional background.  The most general set of covariantized 
gravitational operators consistent with the symmetries of the theory (supersymmetry, 
presence of massless modes) is 

\bqr  
R, \qquad R{1\over\Box}R, \qquad R^2,\qquad  \Box^n R^k \qquad ({\rm mixed~tensors}) 
\fqr 
\bqr  
\ln^{n_1}(\Box)\ldots \ln^{n_m}(\Box) \Box^n R^k 
\fqr 
with the latter terms coming from integrating out massless modes.  The general derivative 
term is explained in detail in \cite{Chalmers1}-\cite{Chalmers16}, and the derivatives are 
in general placed within 
the contractions of the curvature terms.  The duality invariant moduli dependent 
coefficients of the gravitational terms are determined from the ring of functions (within 
the S-duality compliant scattering), 

\bqr 
\prod E_{K3,s_i}^{(q_i,-q_i)} \ , \qquad \sum s_i = s = {3\over 2} + {n\over 2} 
 {(k-4)\over 2}, \qquad \sum q_i = 0 \ , 
\fqr 
and 

\bqr 
{\cal E}_{K3,s}^{(q,-q)} = \sum_{(n^a,m^a)} {1\over (n^a G_{ab} m^b)^{s+q} 
 (n^a G_{ab} m^b)^{s-q}} \ , 
\fqr 
with the sum over the coprime invariants of the toroidal coset metric (the $E_1$ function 
is to be replaced with its regularized version).  The moduli of the space in this 
description have absorbed a factor of the string scale so as to be dimensionless (e.g. 
Wilson lines in toroidal compactifications in the the lattice sum at genus one.)  The 
on-shell supersymmetric completion is required to complete the terms in the generating 
function of the full scattering.  

Next, the coefficients of the terms in the ring of the automorphic functions are described 
iteratively in the derivative expansion; subsequently, the metric on K3 is determined.  

\section{Iterative procedure to coefficients} 

In the iterative procedure to determining the relative coefficients of the ring of 
automorphic coefficients, a series of simple x-space integrals are required; the main 
complication is the tensor algebra and supersymmetric completion.  Once done to a 
given order in derivatives, we can exploit the modular properties of the scattering 
to finding some very non-trivial structure of the Calabi-Yau metrics.  In this section 
we only require the gravitational sector, but we need to eliminate the massive modes 
from the scattering in order to obtain the coefficients, without involving the complications 
of the explicit massive string to accomplish the same, but the matter couplings derived 
on the K3 are not known in this work.) 

To a given order in the genus expansion, via the coupling $\tau_2^{3/2-2g}$ at genus $g$, 
the massive modes begin to contribute at order $\alpha^2$ greate than the massless modes 
(at the four-point order).  This mixes their roles in the derivative expansion, as opposed 
to in the coupling expansion.  This is obviously true at loops $0$ to $2$ and due to 
unitarity true also at higher orders (an analysis is performed in \cite{Chalmers14} and 
\cite{Chalmers16}).  
In order for us to obtain the coefficients of the modular functions in the duality invariant 
scattering without these massive modes, these contributions are to be eliminated through an 
expansion of the K3 massive forms.  

The expansion at derivative order $2(m+4)$ begins with a coefficient at $\alpha^m$.  A 
modular constrained form of  
string coupling expansion at this order in derivatives has the form in the Einstein frame 
\cite{Chalmers16}, 

\bqr 
a_{m,0} \tau_2^{3/2+m/2} + a_{m,1} \tau_2^{-1/2+m/2} + \ldots + 
 \tau_2^{3/2-2g_{\rm max} + m/2} \ , 
\label{couplingexp} 
\fqr 
and maximum genus contribution, $3/2+m/2-2g_{\rm max} = -1/2-m/2$, dictated by the 
expansion of the modular functions.  Furthermore, there are a series of instantonic 
terms coming from the wrapped membranes within the internal space.  The truncation 
of the massive modes has been examined in \cite{Chalmers16} and corresponds to throwing away
half the terms in \rf{couplingexp}.  There are an even number of terms ranging with 
indices of $3/2+m/2$ to $-1/2-m/2$, a total of $1+[m/2]$ terms.  The truncation 
corresponds to dishing the first half terms.  

In the case of the K3 metric the expansion in $\tau_2$ is the same as in \rf{couplingexp}, 
but the coefficients $a_{m,g_{\rm max}}$ are dependent on the moduli non-trivially.  Due 
to the multiplicative nature of the Eisenstein functions on K3, this truncation involves 
first multiplying the various E-functions contributing at a given derivative order $2m$. 

The perturbative terms have integral or half-integral powers in $\tau_2$, and the 
non-perturbative terms have exponential factors; the latter corresponds to wrappings 
of D-branes on the $1$ and $2$-cycles of the K3 space.  The contribution of these 
wrappings are computable via the expansions of the modular functions, similar to 
instantons in ${\cal N}=4$ gauge theory \cite{Chalmers10}.  

Next, the sewing of the derivative terms (with the determined modular coefficients 
$a_{m,g}$) is performed to find the relative coefficients of the modular functions 
generating at each order in derivatives.  This follows via some straightforward 
integrals, however, with some non-trivial tensor components.  The implementation is 
very well suited to be both analytically computed and numerically computed.  

The n-point gravitational vertices derived from the effective action are, 

\bqr 
v_g^{\mu_\sigma,\nu_\sigma} = \prod {\delta\over\delta {\hat g}}_{\mu_j\nu_j} 
 S[g+{\hat g}] \ . 
\fqr 
The fermionic and remaining bosonic vertices are similarly derived, or found by 
supersymmetrization.  The recursive implementation at the four-point order involves 
the sewing relation and the complete set of operators in the derivative expansion.  
This sewing procedure is identical to determining higher loops from lower loops in 
the usual Feynman diagrams; this is clear by examining the low-energy theory of the 
usual loop expansion via a momentum expansion of the graphs.  The sewing generates, 
in momentum space, 

\bqr 
\sum_{L=1}^\infty \int \prod_{j=1}^{L+1} d^dq_j v_g^{2,L+1} \prod_{i=1}^{L+1} 
  \Delta_{\mu_j\nu_j;{\bar\mu}_j{\bar\nu}_j} {\tilde v}_g^{2,L+1} 
\fqr 
\bqr 
+ \sum_{L=1}^\infty \int \prod_{j=1}^{L+1} d^dq_j v_g^{3,L+1} \prod_{i=1}^{L+1} 
  \Delta_{\mu_j\nu_j;{\bar\mu}_j{\bar\nu}_j} {\tilde v}_g^{1,L+1} + {\rm perms} 
 = {1\over 2} v_g^4 \ .    
\fqr 
The complete set of derivatives has been included in the vertices; there are an infinite 
number of terms in $\nu$ that have to b expanded as in $v_4=\sum v_{4,j}$.  The graviton 
propagator is $\Delta_{\mu\nu;\alpha,\beta}$, and the integrals are easiest to evaluate 
by Fourier transforming them to x-space followed by possibly transforming back to x-space.  
The diagram in Figure 1 illustrates the sewing relation.  This iterative procedure, after 
including the fermions, generates all of the required coefficients to any given order in 
derivatives, in the massles mode approximation.  (In order to obtain the full scattering 
the massive modes would have to be included; this is possible by decompactification, 
T-dualizing, and comparing the result with the $d=11$ supergravity coefficients.  In 
principle the contributions of the virtual massive modes to the graviton multiplet scattering 
may be obtained.)   This complets the method of deriving all of the coefficients of the 
gravitational scattering on the Calabi-Yau manifold.

\begin{figure}
\begin{center}
\epsfxsize=12cm
\epsfysize=12cm
\epsfbox{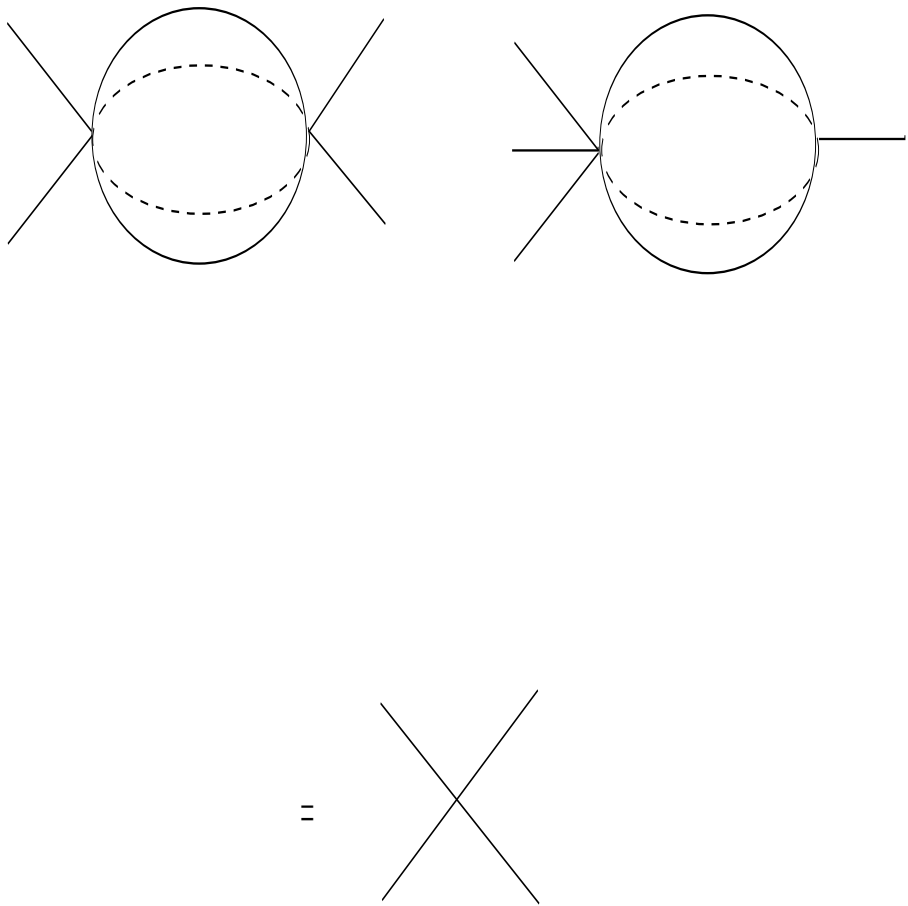}
\end{center}
\caption{The sewing relation illustrated at $4$-point.  Permutations are not included.}
\end{figure}

The integrals required in the sewing procedure are easiest to perform in x-space, 
and the string regulator must be used in order to preserve duality; this is analyzed to 
multi-loops in \cite{Chalmers16}.  The generating function in the classical limit does 
not require 
the supersymmetric completion in the internal legs.  However, as the Eisenstein functions 
receive genus contributions from internal supersymmetric matter, the massless supersymmetric 
completion must be inserted in the loop; an on-shell extension of maximal supergravity is 
developed in \cite{Lightcone}.  The inclusion of the remaining modes, ${\cal N}=4$ 
supergravity and 
${\cal N}=4$ supersymmetric gauge theory complicates the quantum description.  

\section{Determination of the K\"ahler potential} 

The K3, and any Calabi-Yau n-fold, metrics are described locally via the K\"ahler 
potential $g_{i\bar j}=\partial_i\partial_{\bar j}\phi$ (or $g_{i\bar j}=\partial_i 
\partial_{\bar j}\ln\phi$) with a $\mu_k$ moduli dependent scalar $\phi(\mu_k)$.  We 
may substitute this scalar in the full generating function of the graviton scattering 
amplitude in place of the four-dimensional metric.  The general form of the graviton 
scattering function, classical and quantum, to all point orders is

\bqr 
S_{\rm grav}(\phi) = \int d^4x {\cal S}(\phi) \ , 
\fqr 
with 

\bqr 
{\cal S}(\phi)=\sum_{k=1}^\infty {\cal O}_k(\phi) \beta_k \ .  
\fqr 
The coefficients $\beta_k$ are determined to any order via the method given in the 
previous sections.  The operators ${\cal O}_k$ can be complicated due to the 
logarithms and the derivatives; however, only the K\"ahler scalar is involved in 
the expansion.  

With this description we take the condition, 

\bqr 
{\cal S}(\phi) = 0 \ , 
\label{minimization}
\fqr 
which corresponds to minimizing the classical (or quantum action); this is without 
using partial differentiation to enforce the vanishing of the action.  The variation 
is similar to minimizing the scalar potential to find the true vacuum as in the 
Coleman-Weinberg mechanism; this generates an iterative means to obtaining 
the potential in terms of the modular functions $E_s^{(q,-q)}$.  The variation seems 
trivial, but only for the true K\"ahler potential, rather than some general $\phi$, 
will the condition in \rf{minimization} hold.  Due to the specific set of automorphic 
functions pertaining to the K3, and given one metric (and the moduli spanning this 
metric) for any set of functions, the solution should be unique.  

The classical generating function of $\phi$, including the instantons, depends on the 
values of the moduli of the K3 due to the background field expansion of the $R$ term 
about the metric; the terms coming from the multi-derivative curvatures generate an 
equation for $\phi(\mu_k)$.  The quantum terms involving $\phi(\mu_k)$ generate a 
quantum extension for the metrics.  

In general the solution will contain exponentials due to solving \rf{minimization} in 
the presence of logarithms (in the quantum case) and also due to massless singularities 
(in the classical case).  An example series in terms of the K\"ahler scalar is, and for 
simplicitiy we ignore the determinant $\sqrt{g}$, 

\bqr 
{1\over 2} \phi\Box\phi + {1\over 2} g_1 \ln(\phi\Box) \phi\Box\phi+{1\over 2} g_2 
(\phi\Box\phi)^2 + \ldots , 
\fqr 
with $g(\mu_j)$ a sample modular function and $\mu_j$ the moduli.  The $\alpha'$ in the 
expansion sets the dimensional scale in the derivative expansion; its factor scales the 
moduli to be dimensionless (the Eisenstein functions are dimensionless).  Setting this 
function to be zero, and after a field redefinition $\phi=e^\sigma$ makes the equation 
look like a Liouville equation, which is common in the case of Monge-Ampere equations 
in the construction of hyperk\"ahler metrics.  

To higher orders in curvatures there are $\phi^n$ terms with $n$ an arbitrarily large 
integer.  In the case of the classical metric only there are no logarithms, but rather 
an infinite number of these $\phi^n$ terms with many derivatives acting on them.  The 
full solution to the minimization of the action generates the K\"ahler metric, without 
integrating by parts to force the vanishing of the action.  

Because unitarity mixes graphs at different orders, the full loop expansion is 
required to obtain the quantum metric; there is no truncation of the expansion 
that generates the full metric in this approach.  As long as the found potential 
$\phi$ generates a non-singular metric the solution would have to correspond to 
the general metric on K3 simply by symmetries (compact hyperk\"ahler) and 
smoothness; the latter is a consistency check on S-duality and the computed 
expansion.  

\section{Generalization to other Calabi-Yau manifolds} 

The same procedure utilized on the K3 manifolds may be generalized to a variety 
of other compactifications.  First, {\it non-compact} four-dimensional 
manifolds and their respective metrics may be examined with both the string 
scattering and metric information.  This set includes a variety of $4$-dimensional 
hyperk\"ahler metrics such as the A,D, and E series.  

The Calabi-Yau manifolds of complex dimension $3$, $4$, and $5$ may be examined 
through the compactifications $M_{10}\rightarrow T^4\times M_{3-{\rm CY}}$, 
$M_{10}\rightarrow T^2\times M_{4-{\rm CY}}$, and $M_{10}\rightarrow M_{5-{\rm CY}}$.  
These scenarios have $N=16$ in $d=6$, $d=8$, and $d=10$.  A similar procedure 
as described in the previous may be used to compute the metrics on all of the 
compact Calabi-Yau manifolds having a manageable moduli space, such as the toroidal 
ones of projective degenerations; this requirement comes from the requirement 
of explicit modular forms constructable on the manifolds' moduli spaces.  The 
set of CYs also includes the Gepner models.  (Non-compact higher dimensional 
Ricci-flat manifolds are also in principle available, with the moduli spaces 
being the necessary ingredient.)  

Furthermore, the seven dimensional $G_2$ Joyce manifols may be analyzed and 
their metrics computed for classes in which the automorphic functions can 
be computed on the moduli spaces; the latter being useful for M-theory 
compactifications.  The models considered are IIB on $T^3\times M_{G_2}$.  

\section{Curves}  

In this section we examine the exponential terms arising from the expansion 
of the automorphic forms and the coefficients multiplying them.  These 
terms correspond to wrapping of membranes on the internal cycles of the 
K3 (and CY folds), and explicit numbers may be found from the graviton 
scattering.  Both the exponential factors and the rational functions 
multiplying them are of mathematical and physical interest.  The 
factors are cohomological, potentially corresponding to invariants.  

The modular functions have the instanton expansion, 

\bqr  
g_k^{\rm inst} = b_k(G) e^{W(G)} \ , 
\fqr 
for a general configuration of the moduli.  The instanton series are 
computed via the coefficients of the modular functions and is formed 
physically via the wrapping of membranes on the compact cycles.  The 
coefficients $b_k(G)$ of the exponential terms model the counting 
of these wrapped modes in the string coupling expansion, and include 
the moduli of the manifold.  

\section{Summary} 

The derivative expansion of IIB superstring theory on Calabi-Yau spaces 
is examined with the intent of finding detailied metric information of 
the background geometry.  A recipe for computing the metrics on K3 
and higher dimensional Calabi-Yau manifolds, including Joyce manifolds, 
is presented, given information of S-duality and string scattering.  
To any amount of accuracy these Ricci-flat metrics may be computed 
via IIB string theory of M theory amplitudes.  The formalism does 
not have complicated integrals, but rather somewhat complicated 
tensorial algebra at higher orders; as a result, the implementation 
of the recipe is suitable for a computer implementation.  

Furthermore, the metric information found from low orders of derivatives 
in the target space-time theory leads to a natural means of computing numbers 
of holomorphic curves of varying orders and the moduli dependent 
coefficients.  In principle 
this explicit curve and metric information is computable with the use of 
the $T\times S$ modular functions as coefficients in the gravitational scattering.

\vfill\break
 
\end{document}